\documentclass[%
 reprint,
 amsmath,amssymb,
 aps,
]{revtex4-2}

\usepackage{breqn}
\usepackage{amsmath}
\usepackage{color}
\usepackage{graphicx}
\usepackage{dcolumn}
\usepackage{bm}
\usepackage{hyperref}

\hypersetup{
    colorlinks=true,
    linkcolor=blue,
    citecolor=blue,
    filecolor=magenta,      
    urlcolor=blue,
    pdftitle={Overleaf Example},
    pdfpagemode=FullScreen,
    }

\begin{document}

\preprint{APS/123-QED}

\title{The N-Body 2PN Hamiltonian and Numerical Integration of the Equations of Motion}

\author{Felix M. Heinze}
\author{Gerhard Schäfer}
\author{Bernd Brügmann}
\affiliation{Friedrich-Schiller-Universität Jena$,$ Theoretisch Physikalisches Institut$,$ 07743 Jena$,$ Germany}
\date{\today}

\begin{abstract}
To date, the second-order post-Newtonian (2PN) Hamiltonian has been known in closed analytic form only for systems of up to three point masses. In this paper, we present an analytic expression for the general $N$-body 2PN Hamiltonian in the ADM gauge up to a single integral term that, to our knowledge, has no known closed-form analytic solution. We show that the integrals appearing in the 2PN Hamiltonian can be evaluated numerically to machine precision, allowing for cross-validation against analytical results and enabling the full numerical computation of the $N$-body 2PN Hamiltonian. Furthermore, we demonstrate the practical feasibility of the numerical integration of the equations of motion for $N$ bodies at 2PN order using different methods and discuss several strategies for improving computational efficiency.
\end{abstract}

                              
\maketitle


\section{\label{sec:introduction}Introduction}
Understanding the motion of $N$ gravitating bodies is a longstanding problem in astrophysics and celestial mechanics. The two-body problem permits exact analytic solutions in the Newtonian theory. Due to the high symmetry and reducibility to a single effective one-body system, it was possible to develop several formalisms to include relativistic corrections to the two-body problem up to the fifth and partially sixth post-Newtonian (PN) order in the potential contributions, which have proven to be remarkably effective in describing the dynamics and gravitational-wave emission of compact binaries up to the late-inspiral phase \cite{Pati:2000vt, Futamase:2007zz, Blanchet:2013haa, Schafer:2018jfw, Levi:2018nxp, Blumlein:2022qjy}. The PN approximation applies to systems where the involved gravitational inter-body fields are weak and the characteristic velocities are slow compared to the speed of light. In this regime, one can characterize a bound or almost bound system by a small parameter $\epsilon \sim (v/c)^2 \sim GM/(rc^2)$ (where the last relation comes from the virial theorem), allowing Einstein’s field equations to be solved perturbatively as a series expansion in powers of $\epsilon$. The highest power of $\epsilon$ retained in the series expansion defines the order of the post-Newtonian approximation. 

Other approximation schemes—most notably the post-Minkowskian (PM) expansion, gravitational self-force (GSF) theory, and the effective-one-body (EOB) framework—have been successfully used to model relativistic compact-binary dynamics. These approaches, however, are primarily formulated for the two-body problem and do not straightforwardly generalize to general $N>2$ systems, except for the PM expansion \cite{Damour:2009zoi, Damour:2016gwp, Barack:2018yvs, Jones:2022aji}.

In 2005, numerical relativity matured to the point that the solution of the full Einstein equations for the two-body problem became an option \cite{Pretorius2005,Campanelli2005,Baker2005}.
A large number of fully relativistic simulations of compact binaries consisting of black holes, neutron stars, and more exotic objects have been carried out, revealing numerous interesting results, such as strong-gravity precession and spin dynamics, merger recoil kicks, and accurate theoretical predictions for the gravitational waveforms, the ejection of matter, and the emission of neutrinos and electromagnetic radiation during the merger of two compact objects~\cite{Zlochower:2010sn, Baiotti:2016qnr, brugmann2018fundamentals, Shibata:2019wef, bernuzzi2020neutron, Kyutoku:2021icp, pang2023updated}.

For more than two bodies, even the Newtonian equations of motion do not admit general analytic solutions. Such $N$-body systems exhibit highly nonlinear and chaotic behavior, where small variations in the initial conditions can lead to vastly different outcomes. It has been demonstrated that these chaotic properties can be significantly affected by the inclusion of PN corrections \cite{Zwart:2021qxe, DiCintio:2024vgp}. From an astrophysical standpoint, understanding $N$-body dynamics is crucial in many contexts, including the stability of planetary systems, galactic evolution, star cluster dynamics, and cosmic structure formation. 

The broad range of astrophysical applications has driven major advances in PN methods for $N$-body systems, and even numerical-relativity simulations of $N$-body configurations have been performed \cite{Lousto2008, Campanelli2008, Galaviz2010, Bai:2011za, Ficarra2023, Ficarra2024, Bamber2025, Heinze:2025usf, Qiao:2025crp}. Corrections up to the 3.5PN order are implemented in many modern $N$-body simulations \cite{Mikkola:2007ip, Harfst:2008ru, Aarseth:2012eg, Rodriguez:2017pec, Rantala:2022pqs}. However, the complete $N$-body dynamics can currently only be fully included at 1PN order, using the Einstein-Infeld-Hoffmann equations of motion \cite{Einstein:1938yz}, as well as at the 2.5PN and 3.5PN orders, which include the leading and next-to-leading dissipative effects due to gravitational radiation reaction \cite{Konigsdorffer:2003ue}. Even when including these relativistic corrections, one often has to resort to approximations, as the full computations do not scale well and can therefore quickly become too expensive. In most of the larger $N$-body simulations, only pairwise two-body PN terms are considered. \cite{Will:2013cza} argued that incorporating at least some of the occurring non-pairwise couplings in the equations of motion can be crucial for obtaining reliable results. 

As of today, only up to three bodies can be treated consistently at 2PN order. The three-body 2PN Hamiltonian was derived in \cite{Schaefer1987} (with small corrections in the three-body reduction of four-point functions in the potential terms $\propto G^3$ apart from $U^{\mathrm{TT}}$, see \cite{Lousto2008b, Galaviz2011}) and has since been employed in several numerical studies of three-body systems \cite{Lousto2008b, Galaviz2011, Galaviz2011b, Bonetti2016}. Since then, little progress has been made in extending the 2PN Hamiltonian formalism to systems containing more than three bodies, primarily due to the analytical complexity of the involved integrals. A noteworthy work on the extension of the 2PN approximation to $N$-body systems was done in \cite{Chu:2008xm}, where the author derived (up to certain integrals) the effective Lagrangian for the $N$-body problem up to 2PN order in the harmonic gauge / de Donder coordinates using methods from perturbative field theory. Compared to the three-body Hamiltonian, the key missing piece for describing the gravitational dynamics of $N \geq 4$ bodies is the four-point correlation function occurring in the $N$-body Hamiltonian (see Section \ref{sec:pn_formalism}).

Modeling the accurate dynamics of systems of more than three bodies can be relevant for example in binary-binary encounters of stellar-mass objects \cite{Miller2002, Fregeau2004, Zevin2019, Heinze:2025usf}, in mergers of galaxies containing multiple supermassive black-holes \cite{Valtonen1994}, in hierarchical systems consisting of more than three bodies \cite{Hamers2015, Hamers2017, Safarzadeh:2019qkk, Fragione:2019hqt, Vynatheya:2021mgl}, or in the evolution of the dense cores of globular or nuclear star clusters \cite{Breen:2013vla, Fragione:2021nhb, Rantala:2022pqs, Gaete:2024ovu}. Furthermore, it has been shown in previous studies that PN corrections can significantly affect the secular evolution of hierarchical systems, for example by modifying the effects of the Kozai-Lidov mechanism \cite{Will:2014wsa, Yu2020, Lim:2020cvm, Suzuki:2020zbg, Kuntz:2021hhm, Maeda:2023uyx}. Finally, the four-point correlation function occurring in the $N$-body Hamiltonian is also used in describing the dynamics of continuously distributed matter \cite{Kopeikin1985}.

In the present work, we build a bridge between the analytical treatment and promising numerical approaches. We present an explicit analytic expression for the $N$-body 2PN Hamiltonian in the ADM gauge up to a single integral term that, to our knowledge, has no known closed algebraic form. We demonstrate that it is possible to numerically evaluate this integral to machine precision and use this to numerically integrate the equations of motion for $N$ bodies up to 2PN order. 

The paper is structured as follows. In Section \ref{sec:pn_formalism} we provide a summary of the post-Newtonian formalism for $N$ point particles and present the $N$-body 2PN Hamiltonian. Section \ref{sec:pn_integrals} is dedicated to the numerical evaluation and accuracy analysis of the integrals in the 2PN Hamiltonian, including the integral for which we do not have a closed-form analytic expression. In Section \ref{sec:numerical_tests} we present efficient methods for the force and energy computation, as well as the numerical integration of the equations of motion for $N$ bodies at 2PN order. We use these methods in Section \ref{sec:example_systems} to evolve two example four-body systems and to quantify the relative importance of the four-point correlation function. A summary and discussion of our results can be found in Section \ref{sec:summary}. Throughout this paper we use geometric units where $G=c=1$.

\section{The Post-Newtonian Hamiltonian Formalism}
\label{sec:pn_formalism}
In this section, we describe the general post-Newtonian Hamiltonian formalism leading to the complete 2PN Hamiltonian $H = H_{\mathrm{N}} + H_{\mathrm{1PN}} + H_{\mathrm{2PN}}$ for $N$-body systems. We want to note that gravitational radiation first enters at the 2.5PN order, and therefore the systems described by $H$ are conservative. Adding a 2.5PN or even 3.5PN radiation reaction to the Hamiltonian can be easily done for general $N$-body systems (see, e.g.~\cite{Konigsdorffer:2003ue, Bonetti2016}). However, we do not discuss this further here, since the conservative 2PN Hamiltonian will later allow us to test the robustness of our numerical integration methods by evaluating the conservation of energy, linear momentum, and angular momentum. With the Hamiltonian $H$, one can obtain the canonical equations of motion
\begin{equation}
    \dot{x}_a^i=\frac{\partial H}{\partial p_{ai}}, \quad \quad \dot{p}_{ai} = -\frac{\partial H}{\partial x_a^i},
    \label{eq:eom}
\end{equation}
where $x_a^i$ and $p_{ai}$ are the $i$th components of the position and momentum of the $a$th body. \\
\\
In the Newtonian theory, one obtains
\begin{equation}
    H_{\mathrm{N}} = \frac{1}{2} \sum_a \frac{p_a^2}{m_a} - \frac{1}{2} \sum_a \sum_{b \neq a} \frac{m_a m_b}{r_{ab}}
\end{equation}
and the Hamiltonian to first PN order is given by
\begin{equation}
\begin{aligned}
H_{\mathrm{1 P N}}= &-\frac{1}{8} \sum_a m_a \frac{p_a^4}{m_a^4} -\frac{1}{4} \sum_a \sum_{b \neq a} \frac{m_a m_b}{r_{a b}}\Bigg\{6 \frac{p_a^2}{m_a^2}\\
&-7 \frac{\mathbf{p}_a \cdot \mathbf{p}_b}{m_a m_b} -\frac{\left(\mathbf{n}_{a b} \cdot \mathbf{p}_a\right)\left(\mathbf{n}_{a b} \cdot \mathbf{p}_b\right)}{m_a m_b}\Bigg\} \\ &+\frac{1}{2} \sum_a \sum_{b \neq a} \sum_{c \neq a}  \frac{m_a m_b m_c}{r_{a b} r_{a c}},
\end{aligned}
\end{equation}
where $r_{ab} = |\mathbf{x}_a-\mathbf{x}_b|$, $\mathbf{n}_{ab}=(\mathbf{x}_a-\mathbf{x}_b)/r_{ab}$, and $m_a$ is the mass of the $a$th body. Up to this point, the Hamiltonian applies to general systems of $N$ point particles, and using Equations (\ref{eq:eom}), the equations of motion can be written as
\begin{widetext}
\begin{equation}
\begin{aligned}
\dot{\mathbf{x}}_a
&=
\frac{\mathbf{p}_a}{m_a}
-\frac{1}{2}\frac{p_a^2}{m_a^3}\mathbf{p}_a -\frac{1}{2} \sum_{b \neq a} \frac{1}{r_{ab}}
\Bigl(
6\,\frac{m_b}{m_a}\,\mathbf{p}_a
- 7\,\mathbf{p}_b
- (\mathbf{n}_{ab}\!\cdot\!\mathbf{p}_b)\,\mathbf{n}_{ab}
\Bigr),
\end{aligned}
\end{equation}
\begin{equation}
\begin{aligned}
\dot{\mathbf{p}}_a
&=
- \sum_{b \neq a} \frac{m_a m_b}{r_{ab}^2}\,\mathbf{n}_{ab} -\frac{1}{2} \sum_{b \neq a} \frac{1}{r_{ab}^2}
\Bigg[
\Bigl(
3\,\frac{m_b}{m_a}\,\mathbf{p}_a^2
+ 3\,\frac{m_a}{m_b}\,\mathbf{p}_b^2
- 7\,\mathbf{p}_a\!\cdot\!\mathbf{p}_b
- 3(\mathbf{n}_{ab}\!\cdot\!\mathbf{p}_a)(\mathbf{n}_{ab}\!\cdot\!\mathbf{p}_b)
\Bigr)\mathbf{n}_{ab}
\\
&\qquad\qquad\qquad\quad
+ (\mathbf{n}_{ab}\!\cdot\!\mathbf{p}_b)\,\mathbf{p}_a
+ (\mathbf{n}_{ab}\!\cdot\!\mathbf{p}_a)\,\mathbf{p}_b
\Bigg] + \sum_{b \neq a}\sum_{c \neq a}
\frac{m_a m_b m_c}{r_{ab}^2\,r_{ac}}\;\mathbf{n}_{ab}
\;+\;
\sum_{b \neq a}\sum_{c \neq b}
\frac{m_a m_b m_c}{r_{ab}^2\,r_{bc}}\;\mathbf{n}_{ab}.
\end{aligned}
\end{equation}
At the second PN order, things become more difficult, primarily due to the complicated static potential term
\begin{equation}
    U^{\mathrm{TT}} = - \frac{1}{4 \pi} \int d^3x \sum_{i,j,k} \partial_k f_{ij}^{\mathrm{TT}} \ \partial_k f_{ij}^{\mathrm{TT}},
\end{equation}
with
\begin{multline}
    f_{ij}^{\mathrm{TT}} = f_{ij} - \frac{1}{8} \sum_a \sum_{b \neq a} m_a m_b \Bigg[ \delta_{ij} \Bigg( \frac{1}{r_a r_b} - \frac{2}{r_a r_{ab}} + \frac{2}{r_{ab}^2} \mathbf{n}_a \cdot \mathbf{n}_{ab}\Bigg) \\
    + \frac{\partial}{\partial x^i} \frac{\partial}{\partial x^j} \Bigg( \ln(r_a + r_b + r_{ab}) - \frac{r_a}{r_{ab}} + \frac{r_a^2}{2r_{ab}^2} \mathbf{n}_a \cdot \mathbf{n}_{ab}\Bigg) - \frac{4}{r_{ab}^2} (n_a^i n_{ab}^j + n_a^j n_{ab}^i) \Bigg]
\end{multline}
and
\begin{equation}
    f_{ij} = \sum_a \sum_{b \neq a} m_a m_b \frac{\partial}{\partial x_a^i} \frac{\partial}{\partial x_b^j} \ln(r_a + r_b + r_{ab}),
\end{equation}
arising in the 2PN Hamiltonian, where $r_a = |\mathbf{x}-\mathbf{x}_a|$, $\mathbf{n}_a=(\mathbf{x}-\mathbf{x}_a)/r_a$, and $\mathrm{TT}$ indicates the transverse-traceless part of the affixed tensor (see, e.g.~\cite{Ohta:1974pq}). Due to the double sum in $f_{ij}^{\mathrm{TT}}$ there are four sums over particle indices in $U^{\mathrm{TT}}$. One can therefore decompose $U^{\mathrm{TT}}$ as
\begin{equation}
    U^{\mathrm{TT}} = U^{\mathrm{TT}}_{(2)} + U^{\mathrm{TT}}_{(3)} + U^{\mathrm{TT}}_{(4)},
\end{equation}
where $U^{\mathrm{TT}}_{(2)}$ is the two-point correlation function, for which only two particle indices in the sums are distinct, $U^{\mathrm{TT}}_{(3)}$ is the three-point correlation function, for which three particle indices are distinct, and $U^{\mathrm{TT}}_{(4)}$ is the four-point correlation function, for which all four particle indices are distinct. The two-point correlation function has been calculated in, e.g., \cite{Damour:1985mt} for two particles, and for $N$ particles it is given by
\begin{equation}
    U^{\mathrm{TT}}_{(2)} = - \frac{1}{4} \sum_a \sum_{b \neq a} \frac{m_a^2m_b^2}{r_{ab}^3}.
\end{equation}
The three-point correlation function has been calculated in \cite{Schaefer1987} and is given by
\begin{equation}
\begin{aligned}
    U^{\mathrm{TT}}_{(3)} = - \frac{1}{64} \sum_a \sum_{b \neq a} \sum_{c \neq a,b} \frac{m_a^2 m_b m_c}{r_{ab}^3r_{ac}^3r_{bc}} \Big[& 18r_{ab}^2r_{ac}^2 - 60r_{ab}^2r_{bc}^2 - 24r_{ab}^2r_{ac}(r_{ab}+r_{bc}) +60 r_{ab}r_{ac}r_{bc}^2 \\
    &+56 r_{ab}^3r_{bc} - 72 r_{ab}r_{bc}^3 + 35 r_{bc}^4 + 6 r_{ab}^4  \Big].
    \label{eq:utt3}
\end{aligned}
\end{equation}
In systems with $N=2$, only the two-point correlation function $U^{\mathrm{TT}}_{(2)}$ is required in the potential $U^{\mathrm{TT}}$, for $N=3$, both $U^{\mathrm{TT}}_{(2)}$ and $U^{\mathrm{TT}}_{(3)}$ are required, and for systems with $N \geq 4$, one needs the full $U^{\mathrm{TT}}$ potential, consisting of the two-, three-, and four-point correlation functions. At 2PN order, there are no correlation functions beyond the four-point correlation function. To date, there is no complete closed-form analytic expression for $U^{\mathrm{TT}}_{(4)}$, which means that, at 2PN order, the dynamics can presently be described fully analytically only for systems containing up to three bodies. One can show that the four-point correlation function can be written as
\begin{equation}
\begin{aligned}
     U^{\mathrm{TT}}_{(4)} = \frac{1}{4\pi} \sum_a &\sum_{b \neq a} \sum_{c \neq a,b} \sum_{d \neq a,b,c} m_a m_b m_c m_d \sum_{i,j} \frac{\partial^2}{\partial x_c^i \partial x_d^j} \Bigg[ \frac{\partial^2}{\partial x_a^i \partial x_b^j} \int d^3x \frac{\ln(s_{ab})}{r_c r_d} - \frac{1}{8} \int d^3x \frac{\partial_i \partial_j\ln(s_{ab})}{r_c r_d} \\
    &- \frac{\delta_{ij}}{8} \Bigg( \int d^3x \frac{1}{r_a r_b r_c r_d} - \frac{2}{r_{ab}}\int d^3x \frac{1}{r_a r_c r_d} + \frac{2}{r_{ab}^2} \int d^3x \frac{\mathbf{n}_a \cdot \mathbf{n}_{ab}}{r_c r_d }\Bigg) \\
    &+ \frac{1}{8} \Bigg(\frac{1}{r_{ab}}\int d^3x \frac{\partial_i \partial_j r_a}{r_c r_d} - \frac{1}{2r_{ab}^2} \int d^3 x \frac{\partial_i \partial_j (r_a^2 (\mathbf{n}_a \cdot \mathbf{n}_{ab}))}{r_c r_d} \Bigg) + \frac{1}{2 r_{ab}^2} \int d^3 x \frac{n_a^i n_{ab}^j + n_a^j n_{ab}^i}{r_c r_d}\Bigg],
    \label{eq:UTT4_raw}
\end{aligned}
\end{equation}
with $s_{ab}=r_a+r_b+r_{ab}$. Most of the integrals can be solved with quite lengthy calculations involving several integrations by parts and standard identities. For the case of four (or more) bodies in the 2PN approximation (with $a,b,c,d$ distinct), applying all particle derivatives yields integrands that are at most locally of order $\propto 1/r_k^2$ and therefore locally integrable around each particle position, and at least $\propto 1/r^4$ when approaching infinity. Hence, all volume integrals are well-defined, and no regularization is required. In contrast, in the two- and three-body cases, particle indices can coincide and generate stronger singularities, which require treatments under regularization. One of the less trivial results is the solution for the expression containing the integral 
\begin{equation}
    \int d^3x \frac{1}{r_a r_b r_c r_d} = \int d^3x \frac{\Delta \ln(s_{ab})}{r_c r_d} ,
\end{equation}
which also occurs in Equation (C15) of \cite{Pati:2002ux}. In combination with the occurring partial derivatives with respect to the particle positions and the sums over all particle labels, the integral expression can be written as
\begin{equation}
    \sum_a \sum_{b \neq a} \sum_{c \neq a,b} \sum_{d \neq a,b,c}\sum_{i,j} \delta_{ij} \frac{\partial^2}{\partial x_c^i \partial x_d^j} \int d^3x \frac{1}{r_a r_b r_c r_d} = \frac{4 \pi}{3} \sum_a \sum_{b \neq a} \sum_{c \neq a,b} \sum_{d \neq a,b,c} \frac{1}{r_{ad} r_{bd} r_{cd}},
\end{equation}
as given in Equation (A3.1) of \cite{Ohta:1974pq}. In order to obtain this, one can use the identity
\begin{equation}
    (\mathbf{\nabla}_a + \mathbf{\nabla}_b + \mathbf{\nabla}_c) \cdot \mathbf{\nabla}_d \int d^3x \frac{1}{r_a r_b r_c r_d} = - \int d^3x \frac{1}{r_a r_b r_c} \Delta \frac{1}{r_d} = \frac{4 \pi}{r_{ad} r_{bd} r_{cd}}.
\end{equation}
After evaluating all the integral expressions in Equation (\ref{eq:UTT4_raw}) and performing several simplifications, the four-point correlation function can be written as
\begin{equation}
\begin{aligned}
    U^{\mathrm{TT}}_{(4)} = &- \frac{1}{64} \sum_a \sum_{b \neq a} \sum_{c \neq a,b} \sum_{d \neq a,b,c} \frac{m_a m_b m_c m_d}{r_{ab}^3 r_{cd}^3 r_{ad}^3 r_{bc}^3} \Bigg\{ 16 \frac{r_{ab}^3 r_{bc}^3  r_{cd}^2 r_{ad}^2}{r_{bd}} - 24 r_{bc}^3 r_{ab}^2 r_{cd}^2 r_{ad}^2\\
    &- 30 r_{ad}^4 r_{bc}^3 (r_{ad}^2 + r_{bc}^2 - r_{ac}^2 - r_{bd}^2) + r_{ab}^2 (r_{bd}^2-r_{bc}^2-r_{cd}^2) \Bigg(16 \frac{r_{ab} r_{ad}^3 r_{bc}^2}{r_{ac}+r_{bc}+r_{ab}} \\
    &- 8 r_{ad}^3 r_{bc}^2 + r_{ab} r_{cd}^2 (r_{ac}^2-r_{ad}^2-r_{cd}^2) \Bigg) \Bigg\} + \frac{1}{4 \pi} \sum_a \sum_{b \neq a} \sum_{c \neq a,b} \sum_{d \neq a,b,c} m_a m_b m_c m_d \; I^{\mathrm{ln}}_{ab;cd},
    \label{eq:UTT4}
\end{aligned}
\end{equation}
where
\begin{equation}
\begin{aligned}
    I^{\mathrm{ln}}_{ab;cd} &= \sum_{i,j} \frac{\partial^2}{\partial x_a^i \partial x_b^j} \frac{\partial^2}{\partial x_c^i \partial x_d^j}  \int d^3x \frac{\ln(s_{ab})}{r_c r_d} \\
    &= \int d^3x \ \frac{1}{r_c^2r_d^2} \left[ \frac{(\mathbf{n}_c \cdot \mathbf{n}_{ab} - \mathbf{n}_a \cdot \mathbf{n}_c)(\mathbf{n}_d \cdot \mathbf{n}_{ab} + \mathbf{n}_b \cdot \mathbf{n}_d)}{s_{ab}^2} - \frac{\mathbf{n}_c \cdot \mathbf{n}_d-(\mathbf{n}_c \cdot \mathbf{n}_{ab})(\mathbf{n}_d \cdot \mathbf{n}_{ab})}{r_{ab} s_{ab}} \right]
    \label{eq:ln_integral}
\end{aligned}
\end{equation}
\end{widetext}
is the only integral for which we were not able to derive a closed algebraic form. It occurs in the integral part of $U(G^3)$ in Equation (3.19) of \cite{Ohta:1974kp} that is related to one of the two unperformed integrals in Equation (A5.11) of \cite{Ohta:1973je} (the other integral got solved by \cite{Schaefer1987}), and is equivalent to the one in Equation (45) of \cite{Chu:2008xm}. The 3D integral does mathematically exist and is regular only after at least two of the partial differentiations have been performed. We want to note the symmetry properties
\begin{equation}
    I^{\mathrm{ln}}_{ab;cd} = I^{\mathrm{ln}}_{ba;dc} = I^{\mathrm{ln}}_{cd;ab} = I^{\mathrm{ln}}_{dc;ba},
    \label{eq:symmetries}
\end{equation}
where the symmetry $I^{\mathrm{ln}}_{ab;cd} = I^{\mathrm{ln}}_{cd;ab}$ arises because of 
\begin{equation}
    \Delta \ln(s_{ab}) = \frac{1}{r_a r_b}.
\end{equation}
Besides $U^{\mathrm{TT}}$, the 2PN Hamiltonian contains additional static potential terms $\propto G^3$ with quadruple sums over the particle indices, which are for example given in Equations (3.2) and (3.3) of \cite{Ohta:1974pq}. With this, the 2PN part of the $N$-body Hamiltonian can be written down explicitly and is given in Equation \ref{eq:2PN_hamiltonian} of Appendix \ref{sec:appendix}.

\begin{figure*}[htp!]
    \centering
    \includegraphics[width=1\textwidth]{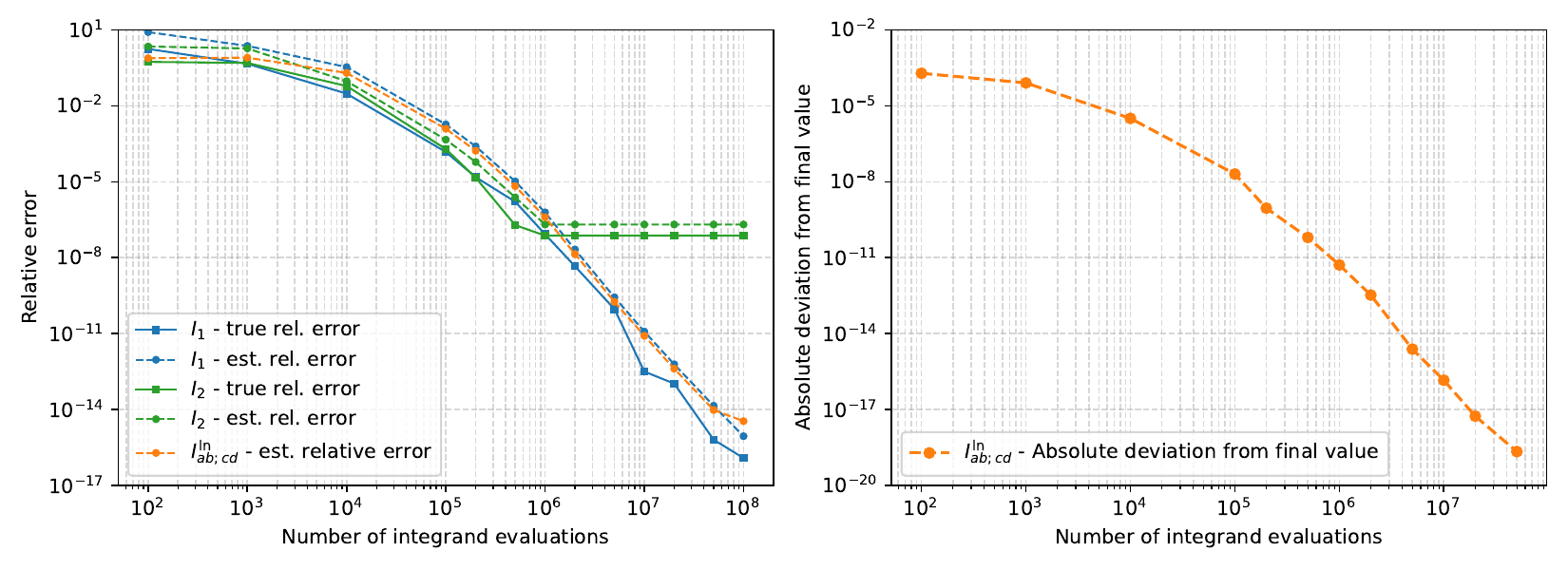}
    \caption{Left: The true and estimated relative errors for the numerical evaluation of three different integrals for a random particle configuration using different numbers of integrand evaluations. Right: The absolute deviation of the value of $I^{\mathrm{ln}}_{ab;cd}$ from the final value with $10^8$ cubature integrand evaluations, indicating convergence to the final result.}
    \label{fig:integral_accuracy}
\end{figure*}

\section{Numerical Evaluation of the 2PN Integrals}
\label{sec:pn_integrals}
In this section, we demonstrate that the integrals appearing in the $N$-body 2PN Hamiltonian can be solved numerically. The integrals occurring in the four-point correlation function (\ref{eq:UTT4_raw}) can be evaluated individually to machine precision, including (\ref{eq:ln_integral}), for which we have no analytic expression.

Part of the numerical challenge is that the integral in question involves an infinite domain, an integrand with two poles in varying locations, and discontinuities due to the $\mathbf{n}$-vectors. For the numerical integration, we used the open-source \textsc{Cuba} library \cite{HAHN200578}. In order to evaluate the 3D integrals, we employed the included routine \textsc{Cuhre}, which is a deterministic cubature algorithm with a globally adaptive subdivision scheme \cite{Berntsen:1991ciy, dcuhre}. At each iteration, the subregion with the highest estimated error is subdivided along the axis corresponding to the largest fourth difference of the integrand. The algorithm is highly effective in moderate dimensions and, within \textsc{Cuba}, is often the only viable option for achieving relative accuracies better than $10^{-3}$, unlike the available Monte Carlo methods. The integrals over $\mathbb{R}^3$ are compactified to the three-dimensional unit cube via a coordinate transformation, without employing any domain decomposition (e.g., isolation of poles). For assessing the accuracy of our numerical methods, we used quadruple precision for all computations.

We evaluated all the integrals occurring in Equation (\ref{eq:UTT4_raw}) as well as the integrals occurring in the two- and three-point correlation functions for various random particle positions and compared the numerical values to the exact results given by their analytic solutions. For the integrands used in this work, we calculated all the occurring partial derivatives analytically, as exemplified in Equation (\ref{eq:ln_integral}). With this, we were able to reproduce the exact values of all the integrals in $U^{\mathrm{TT}}_{(4)}$ with available analytic solutions to machine precision, and the algorithm's estimated errors always served as reliable upper bounds on the true errors. In Figure \ref{fig:integral_accuracy} on the left, we show examples for the true and estimated relative errors for a few selected integrals and for varying numbers of integrand evaluations in the cubature. The integral without a known analytic solution and therefore no true relative error is $I^{\mathrm{ln}}_{ab;cd}$ given in Equation (\ref{eq:ln_integral}). The other integral that was computed numerically to machine precision is
\begin{widetext}
\begin{equation}
\begin{aligned}
    I_1 =\sum_{i,j} \frac{\partial^2}{\partial x_c^i \partial x_d^j}&\int d^3x \frac{\partial_i \partial_j (r_a^2 (\mathbf{n}_a \cdot \mathbf{n}_{ab}))}{r_c r_d} = \frac{2\pi}{r_{ab}} \left( \frac{2}{r_{ac} + r_{ad} + r_{cd}} - \frac{1}{r_{cd}} \right) \left( \frac{r_{ab}^2+r_{ac}^2 - r_{bc}^2}{r_{ac}} + \frac{r_{ab}^2+r_{ad}^2 - r_{bd}^2}{r_{ad}} \right) \\
    & + \frac{\pi}{r_{ab} r_{cd}^3} \Bigg( \frac{(r_{ab}^2 + r_{ad}^2 - r_{bd}^2) (r_{cd}^2 + r_{ad}^2 - r_{ac}^2)}{2 r_{ad}} - \frac{(r_{ab}^2 + r_{ac}^2 - r_{bc}^2) (r_{ad}^2 - r_{ac}^2  -r_{cd}^2)}{2 r_{ac}} \\
    &+ (r_{ad} - r_{ac})(r_{ad}^2+r_{bc}^2-r_{ac}^2-r_{bd}^2) \Bigg),
\end{aligned}
\end{equation}
which occurs in Equation (\ref{eq:UTT4_raw}) for $U^{\mathrm{TT}}_{(4)}$ and which has a known analytic solution. The remaining integral is a two-point integral from Equation (A3a) of \cite{Damour:1985mt} that is similar to $I^{\mathrm{ln}}_{ab;cd}$ but has a known analytic solution given by
\begin{equation}
\begin{aligned}
    I_2 &= \sum_{i,j} \int d^3x \ \partial_i\left( \frac{1}{r_a} \right) \partial_j\left( \frac{1}{r_b} \right) \frac{\partial^2}{\partial x_a^j \partial x_b^i} \ln(s_{ab}) = - \frac{2\pi}{r_{ab}^3}.
\end{aligned}
\end{equation}
\end{widetext}
When evaluating this integral with the same methods used for the other cases, the true relative error levels off at approximately $7.5\times10^{-8}$, while the estimated error remains a safe upper bound. Furthermore, we were able to numerically evaluate all the occurring integrals in the three-point correlation function $U^{\mathrm{TT}}_{(3)}$ and used the results to verify the correctness of the closed-form expression in Equation (\ref{eq:utt3}) from \cite{Schaefer1987}. We therefore conclude that the algorithm’s reported error estimates are reliable for this class of integrals and that our numerical results for $I^{\mathrm{ln}}_{ab;cd}$ can be trusted. This conclusion is further supported by the convergence behavior of the numerical values of $I^{\mathrm{ln}}_{ab;cd}$ with an increasing number of cubature function evaluations, as shown in the right panel of Figure \ref{fig:integral_accuracy}. We want to note that the results shown in Figure \ref{fig:integral_accuracy} are for relatively small-scale configurations. We found that when increasing the particle distances, the estimated errors for $I^{\mathrm{ln}}_{ab;cd}$ become larger at higher numbers of cubature evaluations. This shouldn't be a major problem since in this regime, the integral does not contribute much to the energy and dynamics, as we will show in Section \ref{sec:numerical_tests}.

With the numerical values of $I^{\mathrm{ln}}_{ab;cd}$ we are therefore able to reliably compute $U^{\mathrm{TT}}_{(4)}$ for different particle positions. We found that, independent of the particle positions, $I^{\mathrm{ln}}_{ab;cd}$ largely dominates the value of $U^{\mathrm{TT}}_{(4)}$ and can in general not be neglected.

\section{Numerical Integration of the Equations of Motion}
\label{sec:numerical_tests}

\subsection{Force Computation}

In order to compute the contribution of the integral $I^{\mathrm{ln}}_{ab;cd}$ given by Equation (\ref{eq:ln_integral}) to the total force acting on the particles, we need to compute its gradients with respect to the particle positions. Therefore, in order to solve the equations of motion (\ref{eq:eom}), $I^{\mathrm{ln}}_{ab;cd}$ has to be evaluated many times for each timestep. The total number of integral evaluations per timestep is determined by the ODE integration method, the number of evaluations of $I^{\mathrm{ln}}_{ab;cd}$ required for the computation of its partial derivatives, and it strongly depends on the number of bodies $N$. In practice, all integral evaluations at a given stage of an ODE integration method are independent and can be computed in parallel. \textsc{Cuhre} allows for a vector integrand, which can notably improve the performance, since in many of the computations each integrand differs only a little bit, and some of the common calculations as well as the work for finding a good subdivision of the domain can be shared among all of the integral calculations. In principle, the work can even be shared between subsequent timesteps if the particle positions do not vary significantly, such that the domain subdivision from a previous timestep can still be used. Due to the symmetries given in Equation (\ref{eq:symmetries}), one can additionally save a factor of 4 in the required integral evaluations.

In our setup, we tried to integrate the gradients with respect to the particle coordinates of the integrand of $I^{\mathrm{ln}}_{ab;cd}$, since this would significantly reduce the number of required integral evaluations. Using the analytical expressions of the partial derivatives of the integrand of (\ref{eq:ln_integral}), we only achieved maximum accuracies on the order of a few percent. Using finite differencing inside and outside the integral also introduces a significant numerical error and, if used outside the integral, requires multiple integral evaluations. Instead, we used complex-step differentiation (see e.g.~\cite{ComplexStep}) such that the partial derivative of 
$I^{\mathrm{ln}}_{ab;cd}$ with respect to the $j$th component of the $q$th 
particle position ($q \in \{1, ..., N\}$) is given by 
\begin{equation}
    \frac{\partial I^{\mathrm{ln}}_{ab;cd}}{\partial x_q^j}
    \approx
    \sum_{p \in \{a,b,c,d\}} \delta_{pq}\,
    \frac{\Im\, I^{\mathrm{ln}}_{ab;cd}\bigl(\ldots,
        \mathbf{x}_p + i \varepsilon\,\mathbf{e}_j,
        \ldots\bigr)}{\varepsilon},
    \label{eq:complex_diff}
\end{equation}
where $\mathbf{e}_j$ is the unit vector in the $j$th Cartesian direction, $\Im$ denotes the imaginary part, and $\varepsilon$ is a small real step size that can be set to a small value, e.g., $\varepsilon=10^{-30} M$. We actually did not compute the complex-valued integral $I^{\mathrm{ln}}_{ab;cd}\bigl(\ldots, \mathbf{x}_p + i \varepsilon\,\mathbf{e}_j, \ldots\bigr)$ as suggested by Equation (\ref{eq:complex_diff}), but instead we applied Equation (\ref{eq:complex_diff}) to the integrand of $I^{\mathrm{ln}}_{ab;cd}$, resulting in a single real-valued integral. According to the numerical integrator's estimated error, these integrals can also be evaluated to machine precision. However, there is a problem if $q \in \{c,d\}$, as the singular integrand in Equation (\ref{eq:ln_integral}) can cause severe instabilities in the calculation of the corresponding gradients. One can rewrite the integrand such that it is regular everywhere, but this results in significantly more terms that increase the computation time. Instead, one can also use the symmetry property from Equation (\ref{eq:symmetries}) to obtain
\begin{equation}
    \mathbf{\nabla}_c I^{\mathrm{ln}}_{ab;cd} = \mathbf{\nabla}_c I^{\mathrm{ln}}_{cd;ab}, \quad \quad  \mathbf{\nabla}_d I^{\mathrm{ln}}_{ab;cd} = \mathbf{\nabla}_d I^{\mathrm{ln}}_{cd;ab},
\end{equation}
which also resolves the issue. 

Using the symmetry relations from Equation (\ref{eq:symmetries}), the 3D force evaluation due to $I^{\mathrm{ln}}_{ab;cd}$ for all bodies requires $72\binom{N}{4} \sim O(N^4)$ independent integral evaluations. Each unordered quadruple of bodies $\{a,b,c,d\}$ contributes $4!/4=6$ symmetry-inequivalent permutations, and for each such permutation, one must evaluate the $3 \times 4 = 12$ partial derivatives with respect to the positions of the four particles, giving $6 \times 12 = 72$ scalar integrals per quadruple. These 72 integrals can be computed in parallel for each quadruple using the same domain subdivision. The estimated numerical error of these integrals, as well as comparisons with accurate finite-difference approximations, indicates that this is an accurate, stable, and reliable method for the force evaluations.

\subsection{Energy Computation}
In order to monitor the total energy of the system, one has to evaluate the Hamiltonian at certain timesteps. Here one can proceed in the same way as for the force computation, but one only needs $6\binom{N}{4} \sim O(N^4)$ integral evaluations, i.e., 6 per unordered quadruple of bodies, which can be computed in parallel using the same domain subdivision. For a timestep for which one wants to compute both the forces and the energy, one therefore needs a total of $78\binom{N}{4} \sim O(N^4)$ integral evaluations, and the 78 evaluations per unordered quadruple of bodies can be done with shared intermediate computations and the same domain subdivision.

\subsection{ODE Integration}

The choice of a suitable ODE integration method depends on the intended use case, the desired accuracy, and the available computational resources. In any case, one should aim to reduce the number of expensive numerical evaluations of (\ref{eq:ln_integral}) as much as possible. 

Adaptive high-order methods are an appropriate choice for accurately integrating systems with close encounters. When long-term stability and the conservation of energy, linear momentum, and angular momentum over long time scales are important, symplectic integrators are preferable. In the 2PN $N$-body case, however, explicit symplectic schemes cannot be constructed due to the non-separability of the Hamiltonian into purely kinetic and potential parts. Because of that, one has to resort to (semi-)implicit symplectic integrators (e.g.~\cite{Lubich:2010mj, Seyrich:2012sh, Seyrich:2013jfa}), which can strongly increase the number of integral evaluations due to the fixed-point or Newton iterations needed during the evolution. 

As we will show, for typical situations, $U^{\mathrm{TT}}_{(4)}$ is relatively small compared to the complete Hamiltonian $H$ and becomes significant primarily during close multi-body encounters involving at least four bodies. It could therefore be sufficient to compute $U^{\mathrm{TT}}_{(4)}$ only for close encounters in such subsystems (possibly with a smooth switching function). In situations where the contributions from $U^{\mathrm{TT}}_{(4)}$ are non-negligible, one can split the $N$-body 2PN Hamiltonian from Equation (\ref{eq:2PN_hamiltonian}) into 
\begin{equation}
    H = H_{0} + U^{\mathrm{TT}}_{(4)}(\mathbf{x}_1, ...,\mathbf{x}_N),
\end{equation}
where $H_{0}$ is the part that primarily contributes to the dynamics, and where $U^{\mathrm{TT}}_{(4)}$ contributes much less but is computationally much more expensive, both due to the quadruple sum and due to the numerical evaluation of the integral (\ref{eq:ln_integral}). A suitable approach to integrate the equations of motion is a multiple time stepping method, such as the second-order impulse method (here we primarily follow \cite{Hairer2006})
\begin{equation}
    \Psi_h = \Phi_{h/2}^{(\mathrm{TT},4)} \circ \Big(\Phi_{h/n}^{(0)}\Big)^n \circ \Phi_{h/2}^{(\mathrm{TT},4)},
    \label{eq:multi_time_stepping}
\end{equation}
where $n\geq 1$ is the number of substeps for the method associated with $H_{0}$, $h$ is the outer timestep of the full composition method $\Psi_h$, $h/n$ is the inner (effective) timestep used for the base Hamiltonian $H_{0}$, and $\Phi_{h/n}^{(0)}$ and $\Phi_{h/2}^{(\mathrm{TT},4)}$ are numerical integrators consistent with the Hamiltonians $H_{0}$ and $U^{\mathrm{TT}}_{(4)}$, respectively, and the corresponding equations of motion (\ref{eq:eom}). In the case of $n=1$, this reduces to a second-order Strang splitting
\begin{equation}
    \Psi_h = \Phi_{h/2}^{(\mathrm{TT},4)} \circ \Phi_{h}^{(0)} \circ \Phi_{h/2}^{(\mathrm{TT},4)}.
    \label{eq:strang}
\end{equation}
Since $U^{\mathrm{TT}}_{(4)}$ depends only on the particle positions, the corresponding subflow can be integrated in closed form. The exact subflow $\Phi_{h/2}^{(\mathrm{TT},4)}$ is given by
\begin{equation}
\begin{aligned}
    x_{a,n+1/2}^i &= x_{a,n}^i, 
    \\ 
    p_{ai,n+1/2} &= p_{ai,n} - \frac{h}{2} \,\frac{\partial U^{\mathrm{TT}}_{(4)}}{\partial x_a^i}\big(\mathbf{x}_{1,n},\ldots,\mathbf{x}_{N,n}\big).
\end{aligned}
\end{equation}
This makes $\Phi_{h/2}^{(\mathrm{TT},4)}$ both symmetric and symplectic in exact arithmetic. Since $\partial U^{\mathrm{TT}}_{(4)} / \partial x_a^i$ is computed numerically, this results in the exact kick flow of a numerically defined potential $\tilde{U}^{\mathrm{TT}}_{(4)}=U^{\mathrm{TT}}_{(4)}+\delta U$, i.e., the method is symmetric and symplectic for a modified Hamiltonian $\tilde{H} = H_{0} + \tilde{U}^{\mathrm{TT}}_{(4)}$ up to quadrature and floating-point errors. If one now chooses $\Phi_{h/n}^{(0)}$ to be a symmetric (and/or symplectic) method of order 2 (or higher), then the composition (\ref{eq:multi_time_stepping}) is also symmetric (and/or symplectic) and second order. 

Alternatively, one can use (\ref{eq:strang}) and let $\Phi_{h}^{(0)}$ be an adaptive high-order method completing the timestep $h$ in an adaptive number of substeps. Then the composition (\ref{eq:strang}) generally loses its structure-preserving properties, and while the underlying splitting is second order, in practice the accuracy is primarily governed by the adaptive tolerance during close encounters. Higher-order multiple time stepping methods can be obtained by higher-order splittings, at the cost of additional evaluations of $U^{\mathrm{TT}}_{(4)}$. In both (\ref{eq:multi_time_stepping}) and (\ref{eq:strang}), $\partial U^{\mathrm{TT}}_{(4)}/\partial x_a^i$ effectively only needs to be evaluated once per timestep, since one can reuse $\partial U^{\mathrm{TT}}_{(4)}/\partial x_a^i$ from the end of one timestep for the beginning of the next timestep, because the particle positions do not change during the kick.

\begin{figure*}[htp!]
    \includegraphics[width=0.48\textwidth]{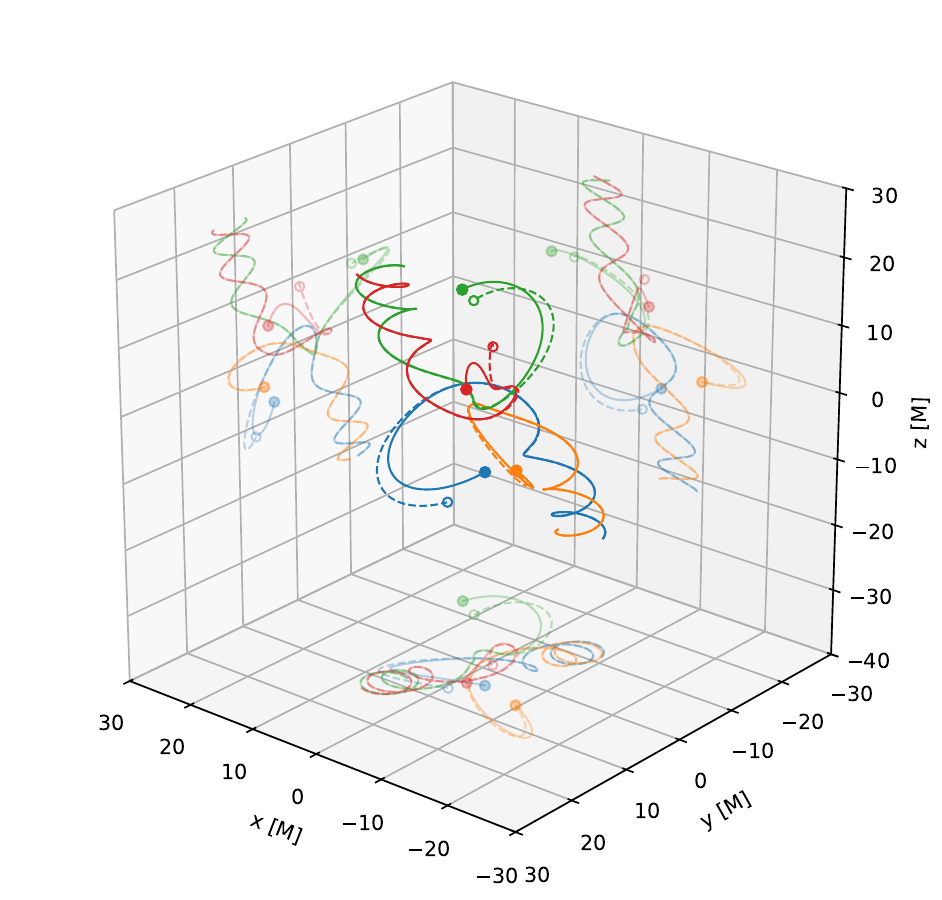}
    \hspace{5pt}
    \includegraphics[width=0.48\textwidth]{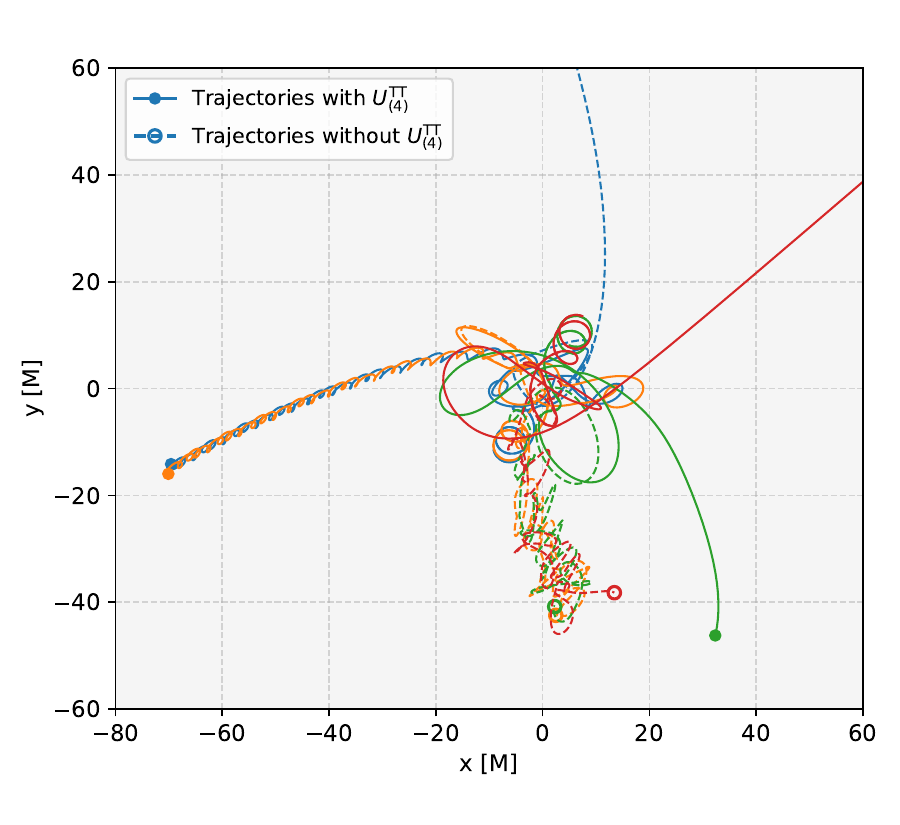}
    \caption{Trajectories of the four bodies in the binary-binary resonance interaction. The solid lines indicate the simulation including $U^{\mathrm{TT}}_{(4)}$, and the dashed lines the one without $U^{\mathrm{TT}}_{(4)}$. The dots indicate the final positions of the bodies (filled including $U^{\mathrm{TT}}_{(4)}$, hollow without $U^{\mathrm{TT}}_{(4)}$). Left: 3D view for the first $650M$ of the evolution with projections onto different planes to give a better sense of the locations in 3D space. Right: $xy$-projection for the full evolution until $t_{\mathrm{final}}=2500M$.}
    \label{fig:bbs_trajectories}
\end{figure*}
\begin{figure*}[htp!]
    \centering
    \includegraphics[width=0.985\textwidth]{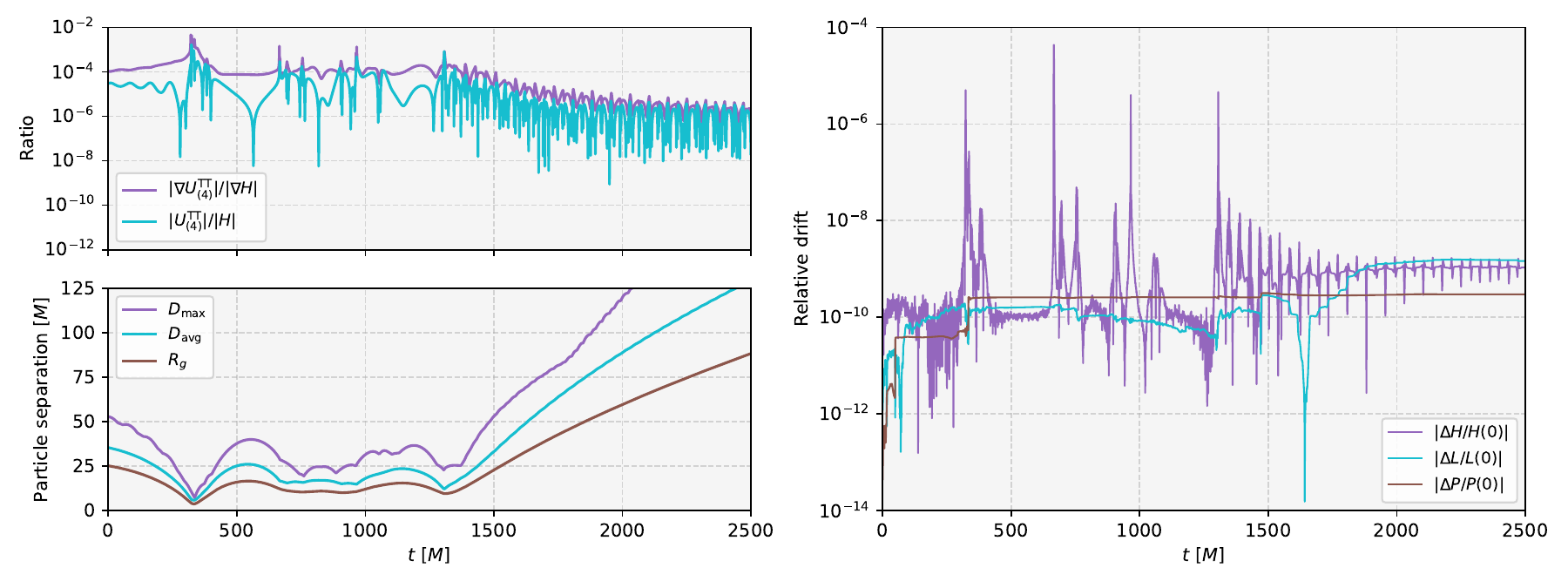}
    \caption{The top left panel shows the contributions of $U^{\mathrm{TT}}_{(4)}$ relative to $H$ as well as the ratio between their gradients $\mathbf{\nabla}U^{\mathrm{TT}}_{(4)}$ and $\mathbf{\nabla}H$ with respect to the particle positions for the close binary-binary encounter. The bottom left panel highlights how the quantities describing the particle separation $D_{\mathrm{max}}$, $D_{\mathrm{avg}}$, and $R_{\mathrm{g}}$ defined in Equations (\ref{eq:D_max})-(\ref{eq:R_g}) vary accordingly. In the panel on the right, the relative errors in the conservation of energy, linear momentum, and angular momentum are plotted.}
    \label{fig:bbs_energies}
\end{figure*}
\section{Evolution of Two Example Systems}
\label{sec:example_systems}
We now present the numerical evolution of two example systems computed with the \textsc{pn-nbody} code, which we developed to integrate few-body dynamics in a range of post-Newtonian (PN) approximations \cite{pn-nbody}. For simplicity, we do not apply any regularization in the examples shown here. Our main purpose is to demonstrate the practical feasibility of such numerical orbit calculations and to quantify the impact of $U^{\mathrm{TT}}_{(4)}$, whose evaluation is dominated by the computationally expensive numerical integral in Equation (\ref{eq:ln_integral}). The initial parameters for these simulations are listed in Table \ref{tb:parameter_values} in Appendix \ref{sec:appendix}.

\subsection{Close Binary-Binary Encounter}
\label{sec:binary-binary}
Our first system is a close binary-binary encounter with a resonance interaction. The system's total mass is $M$, it consists of four equal-mass bodies with $m_k = 0.25M$ ($k=1,2,3,4$), and we evolved the system up to the final time $t_{\mathrm{final}} = 2500 M$. For the numerical ODE integration, we used the Strang splitting (\ref{eq:strang}) with $\Phi_{h}^{(0)}$ being an embedded 5th-order Runge-Kutta method with an adaptive step size to achieve a relative estimated error tolerance of $\mathrm{rtol}=10^{-16}$ in $\Phi_{h}^{(0)}$. For this system, the (outer) step size of $\Psi_h$ is set to $h=0.1M$, and for the numerical evaluation of (\ref{eq:ln_integral}) we used a relative error tolerance of $\varepsilon_{\mathrm{rel}} =10^{-6}$, which corresponds to around $10^6$ integrand evaluations per integral, according to Figure \ref{fig:integral_accuracy}. Our convergence tests have shown that for these values of $h$, $\mathrm{rtol}$, and $\varepsilon_{\mathrm{rel}}$, the trajectories have sufficiently converged. 

Figure \ref{fig:bbs_trajectories} depicts the trajectories of the system. The solid lines show the evolution of the full system, and the dashed lines show the evolution of the system without $U^{\mathrm{TT}}_{(4)}$. On the left, we only show the first $650M$ of the full evolution in a 3D view together with 2D projections. Here, one can clearly see that the solid and dashed trajectories do not visibly differ during the entire approach phase, where the two binaries are further apart. In the strong interaction region, where all four bodies are relatively close, the trajectories start to differ notably, especially the red trajectories, which move in different directions. This already illustrates that the influence of $U^{\mathrm{TT}}_{(4)}$ becomes primarily important when all four bodies are relatively close to each other, and otherwise it is relatively weak. The differing trajectories are, of course, no surprise in this system, as it is highly chaotic in the strong interaction region, such that even tiny perturbations can significantly alter the long-term behavior. This can especially be seen in Figure \ref{fig:bbs_trajectories} on the right, where we plot the $xy$-projection of the full evolution. The outcomes of both simulations are very different due to multiple close encounters. In encounters in which both binaries stay intact and where the four bodies are relatively close only once and for a short duration, $U^{\mathrm{TT}}_{(4)}$ only has a very small effect and does not lead to significant differences. In any case, when averaged over a statistical ensemble of encounters, the 1PN and 2PN terms are generally not expected to alter the final outcomes and are therefore often neglected in the study of stellar encounters (see, e.g.~\cite{Samsing2014}). They do play a much stronger role in the secular evolution of hierarchical systems, which we consider in our second example. 

In Figure \ref{fig:bbs_energies} on the top left, we show the relative contributions of $U^{\mathrm{TT}}_{(4)}$ and its gradient with respect to the particle positions to the full Hamiltonian and its gradient. On the bottom left, we plot the quantities
\begin{equation}
    D_{\mathrm{max}} = \max_{a<b} \ || \mathbf{x}_a - \mathbf{x}_b || ,
    \label{eq:D_max}
\end{equation}
\begin{equation}
    D_{\mathrm{avg}} = \frac{2}{N(N-1)} \sum_{a<b} || \mathbf{x}_a - \mathbf{x}_b ||,
\end{equation}
\begin{equation}
    R_g = \sqrt{\frac{\sum_a m_a (\mathbf{x}_a - \mathbf{R})^2}{\sum_a m_a}} , \quad \mathrm{with} \ \mathbf{R} = \frac{\sum_a m_a \mathbf{x}_a}{\sum_a m_a}.
    \label{eq:R_g}
\end{equation}
One can clearly see that all these quantities anti-correlate with $U^{\mathrm{TT}}_{(4)}$ and its relative contribution to $H$ and its gradient, albeit with stronger oscillations in the contributions of $U^{\mathrm{TT}}_{(4)}$. The oscillations from around $t\approx1300M$ onward are due to the formation of a tight binary (blue and orange trajectories). Our numerical experiments suggest that the average distance of the four bodies $D_{\mathrm{avg}}$ is a good indicator for the relative importance of $U^{\mathrm{TT}}_{(4)}$. The closer the particles are on average, the more important $U^{\mathrm{TT}}_{(4)}$ becomes, with energy and force contributions up to the order of a percent in small-scale configurations. Other numerical experiments have shown that $D_{\mathrm{max}}$ is less suitable for estimating the relative importance of the four-point correlation function. For example, if a body is relatively far away while the other bodies approach one another, the contribution of $U^{\mathrm{TT}}_{(4)}$ increases, but it still remains well below that of a configuration in which all bodies are close. Besides $U^{\mathrm{TT}}_{(4)}$, there are additional terms coming from the other quadruple sums in (\ref{eq:2PN_hamiltonian}), which additionally contribute to the $N$-body Hamiltonian for systems with $N>3$. In this work, we do not aim to quantify their relative contributions to the energy and dynamics, as they can be calculated much more efficiently than $U^{\mathrm{TT}}_{(4)}$.

The plot in Figure \ref{fig:bbs_energies} on the right shows the relative errors in the conservation of energy, linear momentum, and angular momentum. These errors are primarily dominated by the accuracy of the numerical computation of the integral (\ref{eq:ln_integral}). In our example, one can see that, albeit several variations during close encounters, the relative energy errors do not show a strong secular drift on average. Because of the numerically evaluated integral (\ref{eq:ln_integral}), the approximated gradients of $U^{\mathrm{TT}}_{(4)}$ do not satisfy translation and rotation invariance to machine precision. This can introduce small spurious perturbations to the total linear and angular momentum at each kick, most notably when the contribution from $U^{\mathrm{TT}}_{(4)}$ becomes comparatively large. As a result, the deviations in linear and angular momentum accumulate in a diffusive manner, like a random walk, rather than exhibiting a secular drift.

\begin{figure*}[htp!]
    \includegraphics[width=0.48\textwidth]{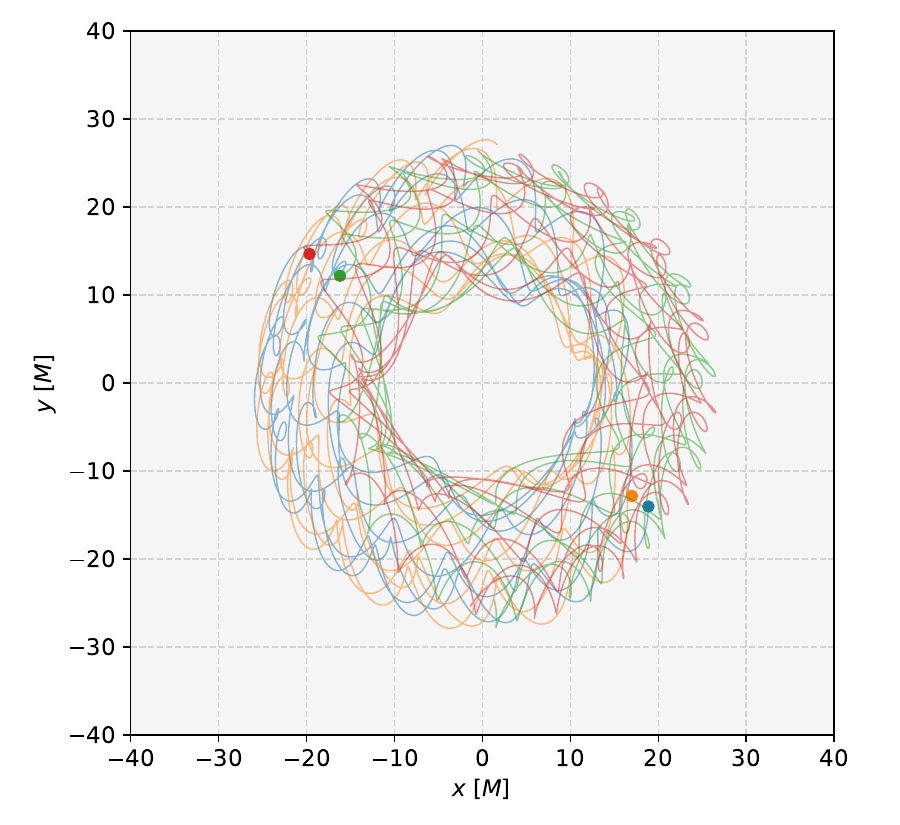}
    \includegraphics[width=0.48\textwidth]{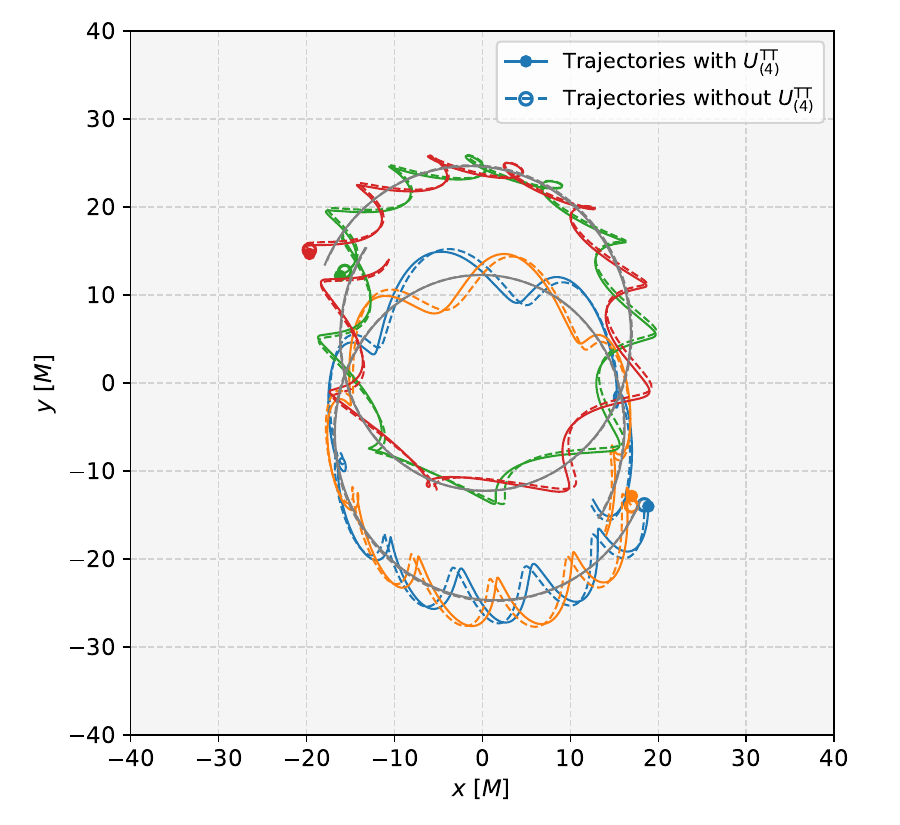}
    \caption{Trajectories of the four bodies in the hierarchical binary-binary system. Left: $xy$-projection of the full trajectories. Right: $xy$-projection of the trajectories of the final outer orbit. The solid lines indicate the simulation including $U^{\mathrm{TT}}_{(4)}$, and the dashed lines the one without $U^{\mathrm{TT}}_{(4)}$. The gray solid and dashed lines indicate the center-of-mass motion of the two binaries. The dots indicate the final positions of the bodies (filled including $U^{\mathrm{TT}}_{(4)}$, hollow without $U^{\mathrm{TT}}_{(4)}$).}
    \label{fig:hierarchical_trajectories}
\end{figure*}
\begin{figure*}[htp!]
    \centering
    \includegraphics[width=1.0\textwidth]{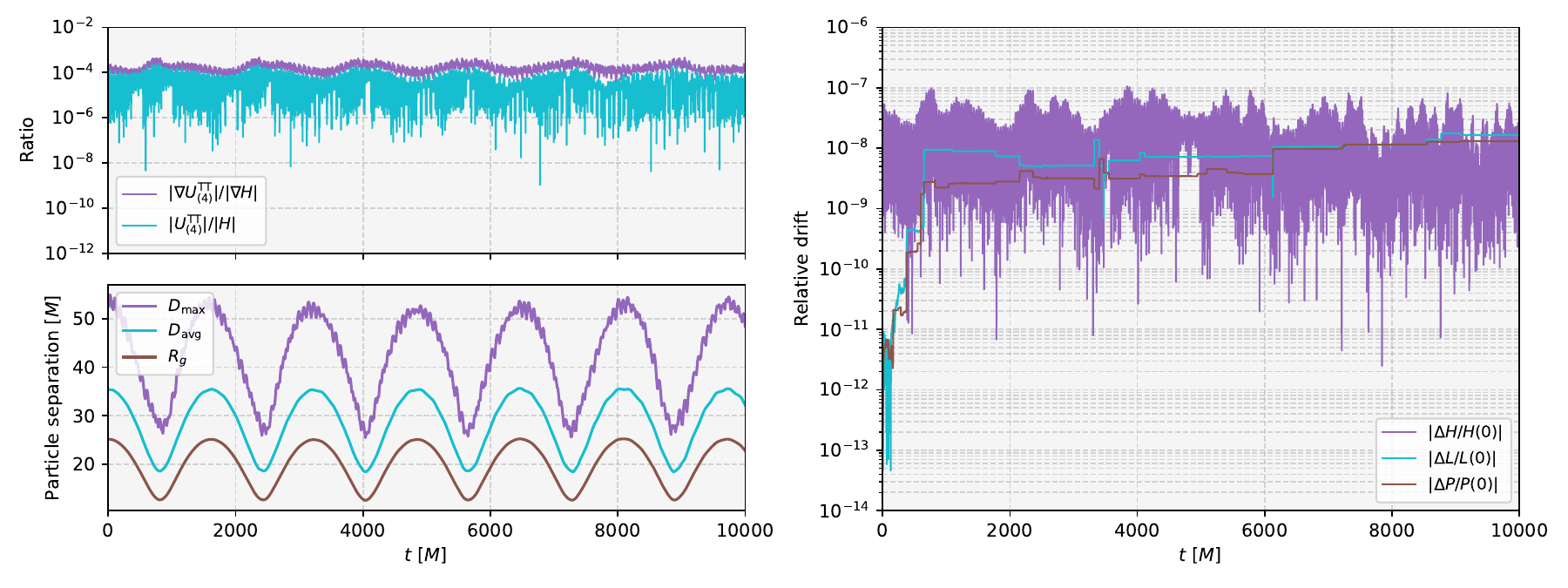}
    \caption{The top left panel shows the contributions of $U^{\mathrm{TT}}_{(4)}$ relative to $H$ as well as the ratio between their gradients $\mathbf{\nabla}U^{\mathrm{TT}}_{(4)}$ and $\mathbf{\nabla}H$ with respect to the particle positions for the hierarchical binary-binary system. The bottom left panel highlights how the quantities describing the particle separation $D_{\mathrm{max}}$, $D_{\mathrm{avg}}$, and $R_{\mathrm{g}}$ defined in Equations (\ref{eq:D_max})-(\ref{eq:R_g}) vary accordingly. In the panel on the right, the relative errors in the conservation of energy, linear momentum, and angular momentum are plotted.}
    \label{fig:hierarchical_energies}
\end{figure*}

\subsection{Hierarchical Binary-Binary System}
\label{sec:hierarchical}
Our second example is a hierarchical system consisting of two similar binaries orbiting each other at a larger distance. The masses of the bodies are the same as in our first example, and we evolved the system up to $t_{\mathrm{final}}=10^4M$. For the numerical integration, we employed the impulse method (\ref{eq:multi_time_stepping}) with $\Phi_{h/n}^{(0)}$ being the implicit midpoint method, which is second order and both symmetric and symplectic. We solved the resulting nonlinear equations using fixed-point iteration with a tolerance of $10^{-12}$. The outer step size is set to $h=1.0M$ and the number of substeps to $n=100$, corresponding to an effective timestep of $h/n = 0.01M$. For the numerical integral evaluations, we chose a relative error tolerance of $\varepsilon_{\mathrm{rel}}=10^{-6}$, which corresponds to around $10^6$ integrand evaluations per integral. Convergence tests with different values of $h$, $n$ and $\varepsilon_{\mathrm{rel}}$ have shown that our obtained results have sufficiently converged.

Figure \ref{fig:hierarchical_trajectories} shows the trajectories of the system. On the left, we depict the trajectories until the final time $t_{\mathrm{final}}=10^4M$ for which $U^{\mathrm{TT}}_{(4)}$ is included. On the right, the last of the six outer orbits is shown, corresponding to the final $1500M$ of the simulation, both with (solid lines) and without $U^{\mathrm{TT}}_{(4)}$ (dashed lines). Here one can see that including $U^{\mathrm{TT}}_{(4)}$ in this system primarily leads to a small phase shift in the individual binaries, while the center-of-mass trajectories (gray lines) and the orientations of the individual binaries remain mostly unchanged. The left plots of Figure \ref{fig:hierarchical_energies} highlight the relative contribution of $U^{\mathrm{TT}}_{(4)}$ to the total energy and the forces in the system for different particle separations. As expected, these relative contributions show oscillations that anti-correlate with the distance of the two binaries, with maximum ratios of around $5\times10^{-3}$ for both the energies and the forces. The plot on the right displays the behavior of the errors in the conservation of the total energy, linear and angular momentum. The energy error shows strong oscillations but remains bounded and does not exhibit a visible secular drift. As in our first example system, the linear and angular momentum errors show a diffusive, random-walk-like behavior, which again results from numerical errors in the evaluation of the integral (\ref{eq:ln_integral}).

\section{Summary and Discussion}
\label{sec:summary}
In this paper, we presented the full $N$-body 2PN Hamiltonian in the ADM gauge (Equation \ref{eq:2PN_hamiltonian}). Compared to the three-body Hamiltonian from \cite{Schaefer1987} and other known results, the new key contribution comes from the four-point correlation function $U^{\mathrm{TT}}_{(4)}$ presented in Equation (\ref{eq:UTT4}). We were able to find a closed-form analytic expression for the integrals occurring in $U^{\mathrm{TT}}_{(4)}$, with the exception of a single integral $I^{\mathrm{ln}}_{ab;cd}$ given in Equation (\ref{eq:ln_integral}), which becomes very difficult to treat analytically in the four-body case with distinct particle labels $a,b,c,d$. Despite the inability to find a complete closed-form expression for $U^{\mathrm{TT}}_{(4)}$, we demonstrated that we are able to reliably compute the remaining integral $I^{\mathrm{ln}}_{ab;cd}$ numerically to machine precision, using the \textsc{Cuba} library. 

Our method enables the computation of the full 2PN Hamiltonian for general $N$-body systems. In addition, it provides a practical, independent validation of existing results for specific integrals obtained through lengthy 
analytic calculations. We tested our numerical method against existing analytic results and used it to verify the correctness of the three-point correlation function $U^{\mathrm{TT}}_{(3)}$ (Equation \ref{eq:utt3}) that was first presented in \cite{Schaefer1987}.

We further demonstrated that it is feasible in practice to use these numerical results to perform accurate numerical orbit calculations. We presented methods to efficiently compute the energy and forces associated with the integral $I^{\mathrm{ln}}_{ab;cd}$, which results in $O(N^4)$ independent and parallelizable integral evaluations per timestep. For the numerical integration of the equations of motion associated with the Hamiltonian (\ref{eq:2PN_hamiltonian}), we used a non-trivial impulse method given by Equation (\ref{eq:multi_time_stepping}), which minimizes the effective number of evaluations of $U^{\mathrm{TT}}_{(4)}$ to one evaluation per outer timestep $h$. This makes it possible to efficiently use both adaptive high-order methods and symplectic methods to reliably evolve few-body systems at 2PN order on short to medium time scales. We demonstrated the practical feasibility of the orbit calculations using two example systems -- a close binary-binary encounter for which we used a high-order adaptive method and a hierarchical binary-binary system for which we employed a method that is both symmetric and symplectic. The errors in the conservation of energy, linear momentum, and angular momentum remain relatively small, they do not show a strong increasing trend, and they are primarily dominated by the accuracy of the numerical computation of $I^{\mathrm{ln}}_{ab;cd}$. Our numerical orbit calculations further revealed that $U^{\mathrm{TT}}_{(4)}$ strongly anti-correlates with the average distance between the individual bodies and that the relative contributions to the energy and dynamics are usually very small (only up to a few percent in very close multi-body configurations). Besides $U^{\mathrm{TT}}_{(4)}$, there are additional terms coming from the other quadruple sums in (\ref{eq:2PN_hamiltonian}), which contribute to the $N$-body Hamiltonian for systems with $N>3$.

In summary, our method enables self-consistent simulations of general $N$-body systems at 2PN order, facilitating systematic studies of 2PN effects and direct comparisons with numerical-relativity predictions for systems with $N > 3$. Additional improvements, such as allowing for efficient orbit calculations for a higher number of bodies could be made by (smoothly) switching $U^{\mathrm{TT}}_{(4)}$ off or modifying the relative error tolerance for the numerical computation of $I^{\mathrm{ln}}_{ab;cd}$ based on the average distance in few-body subsystems. Future work could further focus on the analytic solution or approximation of $I^{\mathrm{ln}}_{ab;cd}$ for specific $N$-body systems, which on the one hand would provide significant theoretical insight, and on the other hand would improve the efficiency of such orbit calculations dramatically. Finally, it would be worthwhile to explore alternative numerical strategies that improve the accuracy and efficiency of evaluating $I^{\mathrm{ln}}_{ab;cd}$.

\section*{Acknowledgments}
FMH has been supported by the Deutsche Forschungsgemeinschaft (DFG) under Grant No. 406116891 within the Research Training Group RTG 2522/1.

The authors gratefully acknowledge the Gauss Centre for Supercomputing e.V. (www.gauss-centre.eu) for providing computing time in project pn36je on the GCS Supercomputer SuperMUC-NG at Leibniz Supercomputing Centre (www.lrz.de). Further computations were performed on the Ara cluster at the Friedrich-Schiller University Jena, funded by the Deutsche Forschungsgemeinschaft (DFG, German  Research Foundation) - 273418955 and 359757177.

\section*{Data Availability}
The \textsc{pn-nbody} code is available at \cite{pn-nbody}. The raw data corresponding to the findings in this manuscript are available from the authors upon request.

\newpage
\appendix
\section{Extra Material}
\label{sec:appendix}
Equation (\ref{eq:2PN_hamiltonian}) is the full expression for the 2PN part of the $N$-body Hamiltonian in the ADM gauge. As described in the main text, $\mathbf{x}_a$ is the position of the $a$th body, $r_{ab} = |\mathbf{x}_a-\mathbf{x}_b|$, $\mathbf{n}_{ab}=(\mathbf{x}_a-\mathbf{x}_b)/r_{ab}$, $\mathbf{p}_a$ is the momentum of the $a$th body, and $m_a$ is the mass of the $a$th body. $I^{\mathrm{ln}}_{ab;cd}$ is given by Equation (\ref{eq:ln_integral}) and it is the only integral expression for which we were not able to find a closed-form analytic expression. Compared to the three-body Hamiltonian, additional contributions for systems with $N>3$ arise in the form of quadruple sums with distinct particle indices.

\begin{table*}[htp!]
\begin{tabular}{c|c|c}
\hline
\multicolumn{1}{|c|}{Example system} & Close binary-binary encounter  & \multicolumn{1}{|c|}{Hierarchical binary-binary system} \\ \hline \hline

\multicolumn{1}{|c|}{\textbf{$m_1/M$}} & 0.25 & \multicolumn{1}{c|}{0.25}  \\ \hline
\multicolumn{1}{|c|}{\textbf{$m_2/M$}} & 0.25 & \multicolumn{1}{c|}{0.25}  \\ \hline
\multicolumn{1}{|c|}{\textbf{$m_3/M$}} & 0.25 & \multicolumn{1}{c|}{0.25}  \\ \hline
\multicolumn{1}{|c|}{\textbf{$m_4/M$}} & 0.25 & \multicolumn{1}{c|}{0.25}  \\ \hline \hline

\multicolumn{1}{|c|}{\textbf{$x_1/M$}} & -9.15192258 & \multicolumn{1}{c|}{-1.65713232} \\ \hline
\multicolumn{1}{|c|}{\textbf{$y_1/M$}} & -11.557844025 & \multicolumn{1}{c|}{22.885323885}  \\ \hline
\multicolumn{1}{|c|}{\textbf{$z_1/M$}} & -21.650635095 & \multicolumn{1}{c|}{-1.319017475}  \\ \hline
\multicolumn{1}{|c|}{\textbf{$x_2/M$}} & -3.34807742 & \multicolumn{1}{c|}{1.65713232} \\ \hline
\multicolumn{1}{|c|}{\textbf{$y_2/M$}} & -10.09279107 & \multicolumn{1}{c|}{27.114676115}  \\ \hline
\multicolumn{1}{|c|}{\textbf{$z_2/M$}} & -21.650635095 & \multicolumn{1}{c|}{1.319017475}  \\ \hline
\multicolumn{1}{|c|}{\textbf{$x_3/M$}} & 5.058522725 & \multicolumn{1}{c|}{1.340201605} \\ \hline
\multicolumn{1}{|c|}{\textbf{$y_3/M$}} & 8.07975174 & \multicolumn{1}{c|}{-24.660050385}  \\ \hline
\multicolumn{1}{|c|}{\textbf{$z_3/M$}} & 21.650635095 & \multicolumn{1}{c|}{2.654438475}  \\ \hline
\multicolumn{1}{|c|}{\textbf{$x_4/M$}} & 7.441477275 & \multicolumn{1}{c|}{-1.340201605} \\ \hline
\multicolumn{1}{|c|}{\textbf{$y_4/M$}} & 13.570883355 & \multicolumn{1}{c|}{-25.339949615}  \\ \hline
\multicolumn{1}{|c|}{\textbf{$z_4/M$}} & 21.650635095 & \multicolumn{1}{c|}{-2.654438475}  \\ \hline \hline

\multicolumn{1}{|c|}{\textbf{$p_{x,1}/M$}} & 0.01061056 & \multicolumn{1}{c|}{-0.013415605} \\ \hline
\multicolumn{1}{|c|}{\textbf{$p_{y,1}/M$}} & -0.04124137 & \multicolumn{1}{c|}{-0.02321345}  \\ \hline
\multicolumn{1}{|c|}{\textbf{$p_{z,1}/M$}} & 0.007501 & \multicolumn{1}{c|}{0.03566595}  \\ \hline
\multicolumn{1}{|c|}{\textbf{$p_{x,2}/M$}} & -0.01061056 & \multicolumn{1}{c|}{-0.016584395} \\ \hline
\multicolumn{1}{|c|}{\textbf{$p_{y,2}/M$}} & 0.04124137 & \multicolumn{1}{c|}{0.02321345}  \\ \hline
\multicolumn{1}{|c|}{\textbf{$p_{z,2}/M$}} & 0.007501 & \multicolumn{1}{c|}{-0.03566595}  \\ \hline
\multicolumn{1}{|c|}{\textbf{$p_{x,3}/M$}} & 0.039141425 & \multicolumn{1}{c|}{0.006824365} \\ \hline
\multicolumn{1}{|c|}{\textbf{$p_{y,3}/M$}} & -0.016774495 & \multicolumn{1}{c|}{-0.040782355}  \\ \hline
\multicolumn{1}{|c|}{\textbf{$p_{z,3}/M$}} & -0.007501 & \multicolumn{1}{c|}{0.00913199}  \\ \hline
\multicolumn{1}{|c|}{\textbf{$p_{x,4}/M$}} & -0.039141425 & \multicolumn{1}{c|}{0.023175635} \\ \hline
\multicolumn{1}{|c|}{\textbf{$p_{y,4}/M$}} & 0.016774495 & \multicolumn{1}{c|}{0.040782355}  \\ \hline
\multicolumn{1}{|c|}{\textbf{$p_{z,4}/M$}} & -0.007501 & \multicolumn{1}{c|}{-0.00913199}  \\ \hline
\end{tabular}
\caption{Masses, initial positions, and initial momenta for the numerical simulations of the close binary-binary encounter and the hierarchical binary-binary system presented in Section \ref{sec:example_systems} of the main text.}
\label{tb:parameter_values}
\end{table*}

Table \ref{tb:parameter_values} lists all the masses, initial positions, and initial momenta for the numerical simulations of the example systems presented in Section \ref{sec:example_systems}.

\begin{widetext}
\begin{equation}
\begin{aligned}
     H_{\mathrm{2PN}}= & \frac{1}{16} \sum_a m_a\left(\frac{p_a^2}{m_a^2}\right)^3+\frac{1}{16} \sum_{a} \sum_{b \neq a} \frac{m_a m_b}{r_{a b}}\left\{10\left(\frac{p_a^2}{m_a^2}\right)^2-11 \frac{p_a^2 p_b^2}{m_a^2 m_b^2}-2 \frac{\left(\mathbf{p}_a \cdot \mathbf{p}_b\right)^2}{m_a^2 m_b^2}\right. \\
    & \left.+10 \frac{p_a^2\left(\mathbf{n}_{a b} \cdot \mathbf{p}_b\right)^2}{m_a^2 m_b^2}-12 \frac{\left(\mathbf{p}_a \cdot \mathbf{p}_b\right)\left(\mathbf{n}_{a b} \cdot \mathbf{p}_a\right)\left(\mathbf{n}_{a b} \cdot \mathbf{p}_b\right)}{m_a^2 m_b^2}-3 \frac{\left(\mathbf{n}_{a b} \cdot \mathbf{p}_a\right)^2\left(\mathbf{n}_{a b} \cdot \mathbf{p}_b\right)^2}{m_a^2 m_b^2}\right\} \\
    & +\frac{1}{8} \sum_{a} \sum_{b \neq a} \sum_{c \neq a} \frac{m_a m_b m_c}{r_{a b} r_{a c}}\Bigg\{18 \frac{p_a^2}{m_a^2}+14 \frac{p_b^2}{m_b^2}-2 \frac{\left(\mathbf{n}_{a b} \cdot \mathbf{p}_b\right)^2}{m_b^2}-50 \frac{\mathbf{p}_a \cdot \mathbf{p}_b}{m_a m_b}+17 \frac{\mathbf{p}_b \cdot \mathbf{p}_c}{m_b m_c} \\
    & -14 \frac{\left(\mathbf{n}_{a b} \cdot \mathbf{p}_a\right)\left(\mathbf{n}_{a b} \cdot \mathbf{p}_b\right)}{m_a m_b}+14 \frac{\left(\mathbf{n}_{a b} \cdot \mathbf{p}_b\right)\left(\mathbf{n}_{a b} \cdot \mathbf{p}_c\right)}{m_b m_c}+\mathbf{n}_{a b} \cdot \mathbf{n}_{a c} \frac{\left(\mathbf{n}_{a b} \cdot \mathbf{p}_b\right)\left(\mathbf{n}_{a c} \cdot \mathbf{p}_c\right)}{m_b m_c}\Bigg\} \\
    & +\frac{1}{8} \sum_{a} \sum_{b \neq a} \sum_{c \neq a} \frac{m_a m_b m_c}{r_{a b}^2}\Bigg\{2 \frac{\left(\mathbf{n}_{a b} \cdot \mathbf{p}_a\right)\left(\mathbf{n}_{a c} \cdot \mathbf{p}_c\right)}{m_a m_c}+2 \frac{\left(\mathbf{n}_{a b} \cdot \mathbf{p}_b\right)\left(\mathbf{n}_{a c} \cdot \mathbf{p}_c\right)}{m_a m_c} +5 \mathbf{n}_{a b} \cdot \mathbf{n}_{a c} \frac{p_c^2}{m_c^2}\\
    &-\mathbf{n}_{a b} \cdot \mathbf{n}_{a c} \frac{\left(\mathbf{n}_{a c} \cdot \mathbf{p}_c\right)^2}{m_c^2}-14 \frac{\left(\mathbf{n}_{a b} \cdot \mathbf{p}_c\right)\left(\mathbf{n}_{a c} \cdot \mathbf{p}_c\right)}{m_c^2}\Bigg\} +\frac{1}{4} \sum_{a} \sum_{b \neq a} \frac{m_a^2 m_b}{r_{a b}^2}\left\{\frac{p_a^2}{m_a^2}+\frac{p_b^2}{m_b^2}-2 \frac{\mathbf{p}_a \cdot \mathbf{p}_b}{m_a m_b}\right\} \\
    & +\frac{1}{2} \sum_{a} \sum_{b \neq a} \sum_{c \neq a, b} \frac{m_a m_b m_c}{\left(r_{a b}+r_{b c}+r_{ac}\right)^2}\left(n_{a b}^i+n_{a c}^i\right)\left(n_{a b}^j+n_{c b}^j\right)\Big\{8 \frac{p_{ai} p_{cj}}{m_a m_c}-16 \frac{p_{aj} p_{ci}}{m_a m_c} \\
    & +3 \frac{p_{ai} p_{bj}}{m_a m_b}+4 \frac{p_{ci} p_{cj}}{m_c^2}+\frac{p_{ai} p_{aj}}{m_a^2}\Big\} \\
    & +\frac{1}{2} \sum_{a} \sum_{b \neq a} \sum_{c \neq a, b} \frac{m_a m_b m_c}{\left(r_{a b}+r_{b c}+r_{c a}\right) r_{a b}}\Bigg\{8 \frac{\mathbf{p}_a \cdot \mathbf{p}_c-\left(\mathbf{n}_{a b} \cdot \mathbf{p}_a\right)\left(\mathbf{n}_{a b} \cdot \mathbf{p}_c\right)}{m_a m_c} \\
    & -3 \frac{\mathbf{p}_a \cdot \mathbf{p}_b-\left(\mathbf{n}_{a b} \cdot \mathbf{p}_a\right)\left(\mathbf{n}_{a b} \cdot \mathbf{p}_b\right)}{m_a m_b} -4 \frac{p_c^2-\left(\mathbf{n}_{a b} \cdot \mathbf{p}_c\right)^2}{m_c^2}-\frac{p_a^2-\left(\mathbf{n}_{a b} \cdot \mathbf{p}_a\right)^2}{m_a^2}\Bigg\} \\
    &- \frac{3}{8} \sum_{a} \sum_{b \neq a} \sum_{c \neq b} \sum_{d \neq c} \frac{m_a m_b m_c m_d}{r_{ab} r_{bc} r_{cd}} - \frac{1}{4} \sum_{a} \sum_{b \neq a} \sum_{c \neq a} \sum_{d \neq a} \frac{m_a m_b m_c m_d}{r_{ab} r_{ac} r_{ad}} \\
    &-\frac{1}{4} \sum_{a} \sum_{b \neq a} \frac{m_a^2 m_b^2}{r_{a b}^3} -\frac{1}{64} \sum_{a} \sum_{b \neq a} \sum_{c \neq a, b} \frac{m_a^2 m_b m_c}{r_{a b}^3 r_{a c}^3 r_{b c}}\Big\{18 r_{a b}^2 r_{a c}^2-60 r_{a b}^2 r_{b c}^2-24 r_{a b}^2 r_{a c}\left(r_{a b}+r_{b c}\right) \\
    &+60 r_{a b} r_{a c} r_{b c}^2 +56 r_{a b}^3 r_{b c} -72 r_{a b} r_{b c}^3 +35 r_{b c}^4+6 r_{a b}^4\Big\} \\
    &- \frac{1}{64} \sum_a \sum_{b \neq a} \sum_{c \neq a,b} \sum_{d \neq a,b,c} \frac{m_a m_b m_c m_d}{r_{ab}^3 r_{cd}^3 r_{ad}^3 r_{bc}^3} \Bigg\{ 16 \frac{r_{ab}^3 r_{bc}^3  r_{cd}^2 r_{ad}^2}{r_{bd}} - 24 r_{bc}^3 r_{ab}^2 r_{cd}^2 r_{ad}^2\\
    &- 30 r_{ad}^4 r_{bc}^3 (r_{ad}^2 + r_{bc}^2 - r_{ac}^2 - r_{bd}^2) + r_{ab}^2 (r_{bd}^2-r_{bc}^2-r_{cd}^2) \Bigg(16 \frac{r_{ab} r_{ad}^3 r_{bc}^2}{r_{ac}+r_{bc}+r_{ab}} \\
    &- 8 r_{ad}^3 r_{bc}^2 + r_{ab} r_{cd}^2 (r_{ac}^2-r_{ad}^2-r_{cd}^2) \Bigg) \Bigg\} + \frac{1}{4 \pi} \sum_a \sum_{b \neq a} \sum_{c \neq a,b} \sum_{d \neq a,b,c} m_a m_b m_c m_d \; I^{\mathrm{ln}}_{ab;cd} \\
\end{aligned}
\label{eq:2PN_hamiltonian}
\end{equation}
\end{widetext}

\interlinepenalty=10000
\clubpenalty=10000
\widowpenalty=10000
\displaywidowpenalty=10000
\bibliography{apssamp} 

\end{document}